\documentclass{conm-p-l}

\usepackage{amssymb,latexsym}

\newtheorem{theorem}{Theorem}[section]
\newtheorem{lemma}[theorem]{Lemma}
\newtheorem{prop}[theorem]{Proposition}
\newtheorem{corr}[theorem]{Corollary}

\newcommand{\cC}{{\mathcal C}} 
\newcommand{\cZ}{{\mathcal Z}} 
\newcommand{\cF}{{\mathcal F}} 
\newcommand{\cV}{{\mathcal V}}
\newcommand{\cW}{{\mathcal W}}
\newcommand{\cP}{{\mathcal P}}
\newcommand{\cH}{{\mathcal H}}
\newcommand{\cE}{{\mathcal E}}
\newcommand{\smalleq}[1]{\mbox{${#1}$}}
\newcommand{\U}{{\rm U}}

\newcommand{\ee}{{\rm e}}
\newcommand{\ii}{{\rm i}}
\newcommand{\dd}{{\rm d}}

\newcommand{\mbf}[1]{\mbox{\boldmath ${#1}$}}

\newcommand{\vx}{{\bf x}}
\newcommand{\vy}{{\bf y}}

\newcommand{\vn}{{\bf n}}
\newcommand{\vm}{{\bf m}}

\newcommand{\vE}{{\bf E}}
\newcommand{\vmu}{\underline{{\mbf \mu}}}

\newcommand{\vzero}{{\mbf 0}}
\newcommand{\Ref}[1]{(\ref{#1})}

\newcommand{\eps}{\varepsilon}
\newcommand{\half}{\mbox{$\frac{1}{2}$}}

\newcommand{\C}{{\mathbb C}}
\newcommand{\Z}{{\mathbb Z}}

\newcommand{\xx}{\stackrel {\scriptscriptstyle
\times}{\scriptscriptstyle \times}}
\newcommand{\eq}{\begin{equation}}
\newcommand{\eqend}{\end{equation}}
\newcommand{\eqa}{\begin{eqnarray}}
\newcommand{\nonueqa}{\begin{eqnarray*}}
\newcommand{\eqaend}{\end{eqnarray}}
\newcommand{\nonueqaend}{\end{eqnarray*}}
\newcommand{\nonu}{\nonumber \nopagebreak \\ \nopagebreak}
\newcommand{\bma}[1]{\begin{array}{#1}}
\newcommand{\ema}{\end{array}}
\newcommand{\bc}{\begin{center}}
\newcommand{\ec}{\end{center}}

\theoremstyle{definition}

\theoremstyle{remark}
\newtheorem{remark}[theorem]{Remark}
\numberwithin{equation}{section}

\begin{document}

\title[CFT and the elliptic Calogero-Sutherland system]{Conformal
field theory and the solution of the (quantum) elliptic
Calogero-Sutherland system}

\author{Edwin Langmann}
\address{Mathematical Physics, KTH Physics, AlbaNova, SE-106 91
Stockholm, Sweden} \email{langmann@theophys.kth.se}
\thanks{Supported by the Swedish Science Research Council~(VR) and the
G\"oran Gustafsson Foundation.}



\copyrightinfo{2005}{American Mathematical Society}

\begin{abstract}
We review the construction of a conformal field theory model which
describes anyons on a circle and at finite temperature, including
previously unpublished results. This anyon model is closely related to
the quantum elliptic Calogero-Sutherland (eCS) system. We describe
this relation and how it has led to an explicit construction of the
eigenvalues and eigenfunctions of the eCS system.
\end{abstract}

\maketitle

\section{Introduction}
The {\it Sutherland system} is defined by the quantum many body
Hamiltonian
\begin{equation}
H_N = -\sum_{j=1}^N \frac{\partial^2}{\partial x_j^2} + \gamma
\sum_{1\leq j<k\leq N} V(x_j-x_k)\label{H}
\end{equation}
with the two body interaction potential
\begin{equation}
V(r) = \frac1{4\sin^2\half r} . \label{V}
\end{equation}
It describes an arbitrary number, $N$, of identical particles moving on a
circle of length $2\pi$, where $x_j\in [-\pi,\pi]$ are the particle
positions, and $\gamma>-1/4$ is the coupling constant. This model is
known to be integrable in the sense that there exist hermitian
differential operators $H_N^{(n)}$ of the form
\begin{equation}
H_N^{(n)} = \sum_{j=1}^N \ii^n \frac{\partial^n}{\partial x_j^n} +
\mbox{lower order terms}
\end{equation}
for $n=1,2,\ldots$, which include the Hamiltonian, $H_N^{(2)}=H_N$,
and which mutually commute, $[H_N^{(n)},H_N^{(n')}]=0$ for all
$n,n'=1,2,\ldots,N$. The explicit eigenvalues and eigenfunctions of
this model are known since a long time \cite{Su}. These eigenfunctions
are essentially equal to the Jack polynomials which are symmetric
functions playing an important role in various areas of mathematics;
see \cite{McD,St}.

As discovered by Calogero \cite{C}, a more
general integrable system can obtained by replacing the interaction
potential above by
\begin{equation}
V(r) = \sum_{m\in\Z} \frac1{4\sin^2\half(r+\ii\beta m)} ,\quad
\beta>0,\label{eV}
\end{equation}
which is essentially the Weierstrass elliptic $\wp$-function with
periods $2\pi$ and $\beta$;\,\footnote{~To be precise: $V(z)=\wp(z)+c_0$
with $c_0 = 1/12 -(1/2)\sum_{m=1}^\infty \sinh^{-2}[(\beta m)/2]$
\cite{WW}.}  see also \cite{OP}.  The Sutherland system is recovered
from this {\it elliptic Calogero-Sutherland (eCS) system} in the
trigonometric limit $\beta\to\infty$. Much less is known about the
eigenfunctions and eigenvalues of the eCS system; see however
\cite{DI,EK,EFK,FV1,FV2,S,T,P} for various recent interesting results
in this direction.

The topic of this paper is a relation of the eCS system to a
particular {\it conformal field theory (CFT)} model describing {\it
anyons on a circle},\,\footnote{~In the physics literature the name
`anyons' is usually used for certain 2+1 dimensional quantum fields,
and we thus stress that our anyons are in 1+1 dimensions.} and the use
of this relation to solve the eCS system.  These anyons are quantum
fields $\phi(x)$ parameterized by a coordinate $x\in [-\pi,\pi]$, and
they are characterized by the following relations,
\begin{equation}
\phi(x)\phi(y) = \ee^{\pm \ii \pi \lambda} \phi(y)\phi(x) \quad \mbox{
for $x \lessgtr y$} , \label{1}
\end{equation} 
where $\lambda>0$ is the so-called {\it statistics
parameter}. Mathematically our anyons can be thought of as generators
of a star algebra represented on some Hilbert space, but they are
somewhat delicate objects: the product of an anyons and its adjoint is
singular as follows,
\begin{equation}
\phi(x)\phi(y)^* = const.\ (x-y)^{-\lambda} [ 1 + O(|x-y|)] \quad
\mbox{ as $x\to y$.}
\label{2}
\end{equation} 
This shows that the anyons are not operators but rather operator
valued distributions.  

Since anyons commute and anticommute for even and odd integers
$\lambda$, respectively, they generalize bosons and fermions to anyon
statistics.\,\footnote{~The name ``{\em any}on'' refers to {\em any}
phase $\ee^{\pm \ii\pi \lambda}$ which appears in the anyon exchange
relations.}  It is important to note that, for odd integers
$\lambda>1$, the anyons are fermion-like but nevertheless different
from fermions: only for $\lambda=1$ is Eq.\ (\ref{2}) consistent with
the relations $\phi(x)\phi(y)^*+\phi(y)^*\phi(x)=\delta(x-y)$
characterizing conventional fermions, and for $\lambda>1$ the
distribution $\delta(x-y)$ is replaced by something more
singular. Anyons with odd integers $\lambda>1$ can be regarded as {\it
composite fermions} \cite{Jain}, and they have been used in effective
theories of the edge excitations in a fractional quantum Hall system
at filling factor $1/\lambda$; see e.g.\ \cite{Wen,PaS}.

One motivation of our work \cite{CL} was an interesting formal
relation between the {\it fractional quantum Hall effect (FQHE)} and
the Sutherland model which we now recall. As is well-known, the
Laughlin wave function \cite{La}
\begin{equation}
\tilde \Psi_0 = \ee^{-\sum_{j=1}^N |z_j|^2/4}
\prod_{1\leq j<k\leq N} (z_j-z_k)^{2m+1}
\end{equation}
is an excellent approximation to the exact ground state of a FQHE
system at filling $1/(2m+1)$, $m=1,2,3\ldots$; $z_j = X_j+\ii Y_j$
with $(X_j,Y_j)$ the electron positions in natural units, and the
system is a disc with radius $R$, $|z_j|\leq R$. On the other hand,
the ground state of the Sutherland system is \cite{Su}
\begin{equation}
\Psi_0 = \ee^{\ii p\sum_{j=1}^N x_j} \prod_{1\leq j<k\leq N}
[\sin\half(x_j-x_k)]^\lambda
\end{equation}
with $p$ the total momentum of the center-of-mass motion ($p$ is
usually set to zero). A simple computation shows that, if one sets
$z_j = R\exp(\ii x_j)$ in the Laughlin wave function $\tilde \Psi_0$,
one obtains the Sutherland groundstate $\Psi_0$ with $\lambda=2m+1$
(up to an irrelevant constant and for some value of $p$). This
suggests to interpret $\Psi_0$ as wave function describing the edge
degrees of freedom of a FQHE system. As mentioned above, anyons have
been used to construct an effective quantum field theory of the edge
excitations of the FQHE, and our original motivation to study anyons
and its relation to the Sutherland system was to get a better
understanding of the FQHE. This has been a useful guide for our work,
even though we were eventually led in quite a different direction. As
mentioned in Section~4.3, we still hope that our results will be
eventually also useful in the context of FQHE physics.

A central result in our work is the following intriguing fact
\cite{CL} (for earlier work see \cite{BPS,MP,Ha,AMOS,Iso,MS,AJL} and
references therein): there exists a self-adjoint operator $\cH$ on the
anyon Hilbert space so that the commutator of this operator with a
product of $N$ anyons is essentially equal to the Sutherland
Hamiltonian acting on this product, i.e.
\begin{equation}
[\cH,\phi(x_1)\cdots \phi_N(x_N)] \Omega = H_N \phi(x_1)\cdots
\phi_N(x_N) \Omega
\end{equation} 
where $\Omega$ is the vacuum state in the anyon Hilbert space, and the
coupling of the Sutherland Hamiltonian $H_N$ in Eq.\ \Ref{H} is
determined by the statistic parameter of the anyons as follows,
\begin{equation}
\gamma = 2\lambda(\lambda-1) . \label{gamma}
\end{equation} 
Since one quantum field operator $\cH$ accounts for arbitrary particle
numbers $N$ of the Sutherland system, we refer to $\cH$ as {\it second
quantization of the Sutherland system}. We will also discuss how to
use this relation to derive explicit formulas for the eigenvalues and
eigenfunctions of the Sutherland system which are equivalent to
Sutherland's solution \cite{EL3}. Different from Sutherland's method,
our approach yields fully explicit formulas for the eigenfunctions,
and it can be generalized also to the elliptic case
\cite{EL1,EL2,EL4}.

In the next Section we describe a rigorous construction of anyons and
the second quantization of the Sutherland system based on CFT methods
due to Graeme Segal \cite{Se} (see also \cite{PS}; in Remark~\ref{r1}
we will also indicate a possible alternative method of proof using
vertex algebras \cite{Kac}). We then explain the second quantization
of the Sutherland systems using these anyons and how to use it to
solve the latter.  In Section~3 we explain our generalization of this
to the elliptic case \cite{EL1,EL2} which has led to an explicit
solution of the eCS model \cite{EL4}. We end with remarks on
alternative proofs, possible extensions, and open questions in
Section~4. In particular, in Section 4.4 we present previously
unpublished formulas for operators which presumably provide a second
quantization for the higher differential operators $H_N^{(n)}$ of the
Sutherland system (for arbitrary $n$).

This paper is a concise but self-contained review, including results
which have not been published before. Many of the proofs are given,
and only for a few technically and/or computationally demanding
arguments we refer to our original papers.

\medskip

\noindent {\bf Acknowledgements.}
I thank the Erwin Schr\"odinger Institute in Vienna for hospitality
where this paper was written. I would like to thank D.\ Serban for
helpful comments.

\medskip

\section{Anyons at zero temperature: trigonometric case}
\subsection{Construction of anyons} 
The starting point of our construction of anyons is the {\it
Heisenberg algebra} which is the algebra with star operation $*$
generated by elements $R$ and $\hat\rho(n)$, $n\in\Z$, obeying the
following relations
\begin{equation}
[\hat\rho(m),\hat\rho(n)]=m\delta_{m,-n},\quad
[R,\hat\rho(n)]=\delta_{n,0}R, 
\end{equation}
and
\begin{equation}
\hat\rho(n)^*=\hat\rho(-n),\quad R^*=R^{-1} 
\end{equation}
for all $m,n\in\Z$, where 
\begin{equation}
Q\equiv \hat\rho(0)
\end{equation}
has the physical interpretation of a {\it charge operator}. This is a
prominent algebra defining a CFT of bosons where $R$ is an intertwiner
between different charge sectors. Mathematically it defines a central
extension of the loop group of smooth maps from the circle to $\U(1)$;
see e.g.\ \cite{PS}.

The standard highest weight representation of this algebra is fully
characterized by a {\it highest weight state} $\Omega$ satisfying
\begin{equation} 
\hat\rho(n)\Omega = 0\quad \forall n\geq 0.
\end{equation}
By standard arguments one then constructs a Hilbert space $\cF$
containing $\Omega$ with an inner product $\langle\cdot,\cdot\rangle$
so that the $*$ is equal to the Hilbert space adjoint; see e.g.\
\cite{PS}. We will refer to $\Omega$ as {\it vacuum} and to $\cF$ as
{\it anyon Hilbert space}.

{\it Regularized anyons} can then by defined as follows,
\begin{equation}
\phi_\eps(x) = \ee^{\sqrt\lambda \sum_{n=1}^\infty \frac1n
\hat\rho(-n) \bar z^n } \ee^{-\ii \lambda Q x/2} R\ee^{-\ii \lambda Q
x/2} \ee^{ -\sqrt\lambda \sum_{n=1}^\infty \frac1n \hat\rho(n) z^n },
\end{equation}
where 
\begin{equation}
z\equiv \ee^{\ii x-\eps},\quad \bar z \equiv \ee^{-\ii
x-\eps} , 
\end{equation}
and $\eps>0$ is a regularization parameter \cite{CL}.  To see that this
is well-defined, we note that $\phi_\eps(x)$ is proportional to the
following unitary operator,
\begin{equation}
V_\eps(x) = R \exp\Bigr(-\ii \lambda Qx - \sqrt{\lambda} \sum_{n\neq
0} \frac1n \hat\rho(n) \ee^{\ii nx - |n|\eps}\Bigr) .
\end{equation} 
Indeed, defining normal ordering $\xx \cdots \xx$ as usual (see e.g.\
\cite{CL}) we have
\begin{equation}
\phi_\eps(x) =  \, \xx V_\eps(x) \xx , 
\end{equation}
and by a straightforward computation using the Hausdorff formula
\begin{equation} 
\ee^A \ee^B = \ee^{c/2} \ee^{A+B} = \ee^{c} \ee^B \ee^A\quad \mbox{ if
$[A,B]=c$ with $c\in\C$,}
\end{equation} 
one finds that $\phi_\eps(x)$ is proportional to $V_\eps(x)$ with a
proportionality constant diverging as $\eps^{-\lambda/2}$ as
$\eps\downarrow 0$. The anyons can then be defined as
\begin{equation}
\phi(x) \equiv \lim_{\eps\downarrow 0} \phi_\eps(x) 
\end{equation}
where the normal ordering ensures that this limit exists.  To check
that these anyons obey the correct exchange relations in Eq.\ \Ref{1}
we again use the Hausdorff relation to compute
$$ \phi_\eps(x) \phi_{\eps'}(y) = \ee^{(\cdots)}
\phi_{\eps'}(y)\phi_\eps(x), 
$$ and we obtain
$$ (\cdots) = -\lambda \Bigl( \ii[x-y] + \sum_{n\neq 0} \frac1n
\ee^{\ii n(x-y) - |n|(\eps+\eps')}\Bigr) \to - \lambda \ii \pi \,
{\rm sign}(x-y) 
$$ as $\eps,\eps'\to 0$.  To show that the anyons are well-defined one
can compute the {\it anyon correlation functions}, i.e., the vacuum
expectation values of products of anyons, and convince oneselves that
they remain well-defined as all regularizations are removed.  These
correlation functions are also of main physical interest.

\begin{prop}
\label{prop1}
The regularized anyons are bounded Hilbert space operators, and
they have well-defined limits $\eps\downarrow 0$ so that all anyon
correlations are well-defined. Denoting products of anyons as
\begin{equation}
\Phi_N(\vx) \equiv \phi(x_1)\phi(x_2)\cdots \phi(x_N), \label{Phi} 
\end{equation}
the non-zero anyon correlation functions are as follows,
\begin{equation}
\langle \Omega, \Phi_N(\vx)^*R^{N-M} \Phi_M(\vy) \Omega\rangle =
\ee^{\ii \half \lambda(N-M)(X+Y) } F_{N,M}(\vx;\vy)\label{vev}
\end{equation} 
for $N,M=0,1,2,\ldots$, where
\begin{equation}
F_{N,M}(\vx;\vy)\equiv 
\frac{\prod_{1\leq j<k\leq N}
\theta(x_j-x_k)^\lambda \prod_{1\leq j<k\leq M}
\theta(y_k-y_j)^\lambda}{\prod_{j,k=1}^N \theta(x_j-y_k)^\lambda}
\label{FNM}
\end{equation}
with
\begin{equation}
\theta(r) \equiv \sin\half r ,\label{tet}
\end{equation}
and
\begin{equation}
X\equiv \sum_{j=1}^N x_j,\quad Y\equiv\sum_{j=1}^M y_j.
\end{equation}
\end{prop}

(The proof is by straightforward computations using the defining
relations of the Heisenberg algebra and the Hausdorff formula; see
\cite{EL2,EL6} for details.)

It is worth noting that the factor $R^{N-M}$ is inserted on the
r.h.s.\ in Eq.\ \Ref{vev} to get a nonzero result (since
$\langle\Omega,A\Omega\rangle = 0$ unless $A$ commutes with $Q$). In
particular,
\begin{equation}
\langle\Omega, \phi(x)^*\phi(y) \Omega\rangle = \theta(x-y)^{-\lambda}, 
\label{2pt}
\end{equation}
consistent with Eq.\ \Ref{2}.

\begin{remark}
\label{r1}
Our regularization has the advantage of producing well-defined
operators so that multiplication of anyons becomes unambiguous, but
the price we pay is that our regularized anyons are not analytic. An
alternative regularization is to analytically continue the anyons to
the complex region outside the unit circle and define
\begin{equation}
\phi(x) = \, \xx\! R \exp\Bigr(-\ii \lambda Qx -
\sqrt{\lambda}\sum_{n\neq 0} \frac1n \hat\rho(n) \ee^{\ii nx} \Bigr)
\!\xx , \quad \Im(x) =\eps >0 .
\label{anyon}
\end{equation} 
These regularized anyons are not operators but only sesquilinar forms,
however, working with them makes computations somewhat simpler. For
example, Eq.\ \Ref{vev} remains true for these regularized anyons as
it stands (for appropriate complex $x_j,y_j$), which is not the case
for our regularization. Moreover, there exists a mathematical
formulation of CFT which allows to give a rigorous meaning to these
sesquilinar forms as generators of a so-called vertex algebra; see
e.g.\ \cite{Kac}.

For simplicity we will mostly ignore this technical issue of
regularization in the following discussion.
\end{remark}

\subsection{Second quantization of the Sutherland model} 
One natural choice for an anyon Hamiltonian is
\begin{equation}
H_0 = \frac{\lambda}2 Q^2 + \sum_{n=1}^\infty
\hat\rho(-n)\hat\rho(n),\label{aH}
\end{equation}
and it satisfies
\begin{equation}
[H_0,\phi(x)] = \ii \frac{\partial\phi(x)}{\partial x}.\label{H0}
\end{equation}
We note in passing that this model is a CFT since $H_0$ is equal to
the zero mode $L_0$ of a $c=1$ representation of a Virasoro algebra
existing in this model (see e.g.\ \cite{Kac,PS}): the generators
$L_n$ are the Fourier modes of $\xx\rho(x)^2 \xx/2$ with
\begin{equation}
\rho(x) = \sqrt{\lambda} Q + \sum_{n\neq 0} \hat\rho(n) \ee^{\ii nx}
\end{equation} 
the boson field in position space. Eq.\ \Ref{H0} suggests to interpret
$H_0$ as second quantization of (minus the) the momentum operator
$\ii\partial/\partial x$. It is interesting to try to find another
anyon Hamiltonian which corresponds to a second quantization of the
second derivative $-\partial^2/\partial x^2$. By straightforward
computations one finds that the operator
\begin{equation}
\cH = \sqrt{\lambda} W^3 -\frac{3\lambda-2}{12}Q + (1-\lambda) \cC
\label{cH}
\end{equation}
with
\begin{equation}
W^3 = \frac13 \int_{-\pi}^\pi \frac{\dd x}{2\pi}\, \xx\! \rho(x)^3
\!\xx,\qquad \cC = \sum_{n=1}^\infty n \hat\rho(-n)\hat\rho(n)
\label{W3cC}
\end{equation}
obeys the relations
\begin{equation}
[\cH,\phi(x)] = -\frac{\partial^2}{\partial x^2}\phi(x) + \cdots
\label{H1}
\end{equation}
where the dots stand for terms which annihilate the vacuum,
$(\cdots)\Omega=0$. This seems to be the closest one can get to a
second quantization of $-\partial^2/\partial x^2$. While $W^3$ above
is a local operator, it is interesting to note that $\cC$ is
non-local: in position space it can be written in the following
suggestive form
\begin{equation}
\cC = -\lim_{\eps\downarrow 0} \int \frac{\dd x}{2\pi} \int \frac{\dd
y}{2\pi} \, \frac{\rho(x)\rho(y)}{4\sin^2\half(x-y+\ii
\eps)} .
\end{equation}

It is interesting to compute the commutator of $\cH$ with a product of
anyon operators. By a straightforward but tedious computation one
finds the relations in Eq.\ \Ref{2nd}.  We summarize these results as
follows.

\begin{prop} 
\label{prop2}
There exist mutually commuting self-adjoint operators $Q,H_0$ and
$\cH$ on the anyon Hilbert space with the following relations,
\begin{eqnarray}
{[}Q,\Phi_N(\vx){]} &=& N \Phi_N(\vx),\\
{[}H_0,\Phi_N(\vx){]} &=& \sum_{j=1}^N \ii \frac{\partial}{\partial
x_j}\Phi_N(\vx), \\ {[}\cH,\Phi_N(\vx){]}\Omega &=& \Bigl(
- \sum_{j=1}^N \frac{\partial^2}{\partial x_j^2} + \sum_{j<k}
2\lambda(\lambda-1) V(x_j-x_k) \Bigr) \Phi_N(\vx)\Omega,\label{2nd}
\end{eqnarray}
with $\Phi_N(\vx)$ in Eq.\ \Ref{Phi} and $V(r)$ in Eq.\ \Ref{V}.
\end{prop} 

(More details and full proofs can be found in \cite{CL}.)

The relation of main interest for us is in Eq.\ \Ref{2nd}, but we
formulated this proposition so as to suggest a more general result: as
discussed in more detail Section~4.4, we believe that there exists a
second quantization of all the commuting differential operators
$H_N^{(n)}$ of the Sutherland system mentioned in the introduction
also for $n>2$, and we also have a specific conjecture for these
operators.

\subsection{Solution of the Sutherland system}
Proposition \ref{prop2} provides a nice method to construct
eigenfunctions of the Sutherland system.

\begin{corr} 
\label{corr1}
Let $\eta$ be an eigenstate of the CFT operator $\cH$ with eigenvalue
$\cE$, $\cH\eta=\cE\eta$, and with charge $N$, $Q\eta=N\eta$. Then the
inner product of $\eta$ with the state $\Phi_N(\vx)\Omega$,
\begin{equation}
\Psi_{\eta}(\vx) \equiv \langle \Omega, \Phi(\vx)^* \eta \rangle,
\end{equation}
is an eigenfunction of the Sutherland Hamiltonian $H_N$ in Eqs.\
\Ref{H}, \Ref{V}, \Ref{gamma} with the same eigenvalue $\cE$,
\begin{equation}
H_N\Psi_{\eta}(\vx) =\cE \Psi_{\eta}(\vx) .\label{HE}
\end{equation}
\end{corr} 

\begin{proof}
We compute $\langle \Omega,\Phi_{\eta}(\vx)^* \cH \eta\rangle= \langle
\cH\Phi_{\eta}(\vx)\Omega,\eta\rangle$ in two different ways: firstly,
using Eq.\ \Ref{2nd} and the highest weight condition
\begin{equation}
\cH\Omega=0\label{hw}
\end{equation}
yields the l.h.s.\ of Eq.\ \Ref{HE}, and secondly using
$\cH\eta=\cE\eta$ gives the r.h.s.
\end{proof} 

Remarkably, Eq.\ \Ref{2nd} can also be used to construct enough
eigenstates $\eta$ of $\cH$ to recover all the eigenfunctions of
$H_N$. The idea is to take the Fourier transform of Eq.\
\Ref{2nd}. For that one has to remember an important detail: the
anyons are not periodic in general but obey
\begin{equation}
\phi(x+2\pi) = \ee^{-\ii \lambda Q \pi}\phi(x) \ee^{-\ii \lambda Q
\pi}
\end{equation} 
with $Q$ the charge operator obeying
$[Q,\phi(x)]=\phi(x)$ and $Q\Omega=0$. Thus we need to
remove the non-periodic factor from the anyons before we can take the
Fourier transform,
\begin{equation}
\hat\phi(n) = \int_{-\pi}^\pi\frac{\dd x}{2\pi} \ee^{\ii
\lambda Q x /2} \phi(x) \ee^{\ii \lambda Q x/2}\ee^{\ii
nx},\quad n\in\Z.
\end{equation}
A straightforward computation then yields the following result.

\begin{prop}
\label{prop3}
The following product of Fourier transformed anyons,
\begin{equation}
\hat\Phi_N(\vn)\equiv \hat\phi(n_1)\hat\phi(n_2)\cdots
\hat\phi(n_N),\quad \vn=(n_1,\ldots,n_N)\in\Z^N,
\end{equation}
obeys the relation
\begin{equation}
[\cH,\hat\Phi_{N}(\vn)]\Omega = \cE_0(\vn) \hat\Phi_{N}(\vn)\Omega -
\gamma \sum_{1\leq j<k\leq N}\sum_{\nu=1}^\infty \nu \hat\Phi_{N}(\vn
+ \nu \vE_{jk})\Omega \label{Heta}
\end{equation}
with
\begin{equation}
\cE_0(\vn) = \sum_{j=1}^N \tilde n_j^2,\quad \tilde n_j = n_j +
\half \lambda (2N+1-2j),\label{cE0}
\end{equation}
and
\begin{equation}
(\vE_{jk})_\ell \equiv \delta_{j\ell}-\delta_{k\ell} ,\quad
j,k,\ell=1,2,\ldots N.
\end{equation}
\end{prop} 

\begin{remark}
As will be explained below, the $\cE_0(\vn)$ in Eq.\ \Ref{cE0} are
equal to the eigenvalues of the Sutherland Hamiltonian. As is
well-known, they are remarkably similar to the eigenvalues of the
non-interacting case: the Sutherland interaction only changes the
momenta $n_j$ to the so-called {\em pseudo-momenta} $\tilde n_j$. It
is important to note the following arbitrariness in the definition of
the pseudo-momenta: if $\Psi(\vx)$ is an eigenfunction of the
Sutherland Hamiltonian $H_N$ with eigenvalue $\cE_0(\vn)=\sum_{j=1}^N
\tilde n_j^2$, then
\begin{equation}
\tilde \Psi(\vx) = \ee^{-\ii p\sum_{j=1}^N x_j} \Psi(\vx)
\end{equation}
is also an eigenfunction with eigenvalue
\begin{equation}
\cE_0'(\vn) = \sum_{j=1}^N (\tilde n_j')^2,\quad \tilde n_j' =
\tilde n_j^{\phantom '} - p.
\end{equation}
Thus the pseudo-momenta can be shifted by changing the center-of-mass
motion which is, of course, a trivial change.  Our definition of
pseudo-momenta here differs from the one in \cite{Su,EL1,EL2} by the
constant $p=N\lambda/2$. As we will see, this later choice puts the
center-of-mass to rest; see Eq.\ \Ref{to_rest} below.
\end{remark}

\begin{proof}
We observe that
\begin{equation}
 \hat\Phi_{N}(\vn)\Omega = \int_{[-\pi,\pi]^N} \frac{\dd^Nx}{(2\pi)^N}
 \, \Phi_N(\vx) \ee^{\ii \tilde \vn\cdot \vx} \Omega
\end{equation}
with $\tilde \vn=(\tilde n_1,\ldots,\tilde n_N)$, where the shifts in
the Fourier modes $\tilde n_j$ come from the factors $\exp(\ii \lambda
Q x_j/2)$ introduced to compensate the non-periodicity of the
anyons. Eq.\ \Ref{Heta} then follows if we recall Eq.\ \Ref{H}, with
the first term on the r.h.s.\ coming from the derivative terms in
$H_N$ and partial integrations, and the second from the interaction
using
\begin{equation}
\frac1{4\sin^2\half(x_j-x_k + \ii \eps)} = -\sum_{\nu=1}^\infty \nu
\ee^{\ii \nu(x_j-x_k+\ii \eps)}
\end{equation} 
with $\eps\downarrow 0$ our regularization parameter. 
\end{proof}

The proof shows that the regularization of the anyons is not just a
technicality but crucial to get correct results.

A simple computation then yields the solution of the Sutherland system
in terms of anyon correlation functions.

\begin{theorem} 
\label{thm1}
The CFT operator $\cH$ has eigenstates $\eta_N(\vn)$ with
corresponding eigenvalues $\cE_0(\vn)$ given in Eq.\ \Ref{cE0}
provided that
\begin{equation}
n_1\geq n_2\geq \ldots \geq n_N. \label{vn}
\end{equation} 
These eigenstates are
\begin{equation}
\eta_N(\vn) = \sum_{\vm} \alpha_{\vn}(\vm)
\hat\Phi_N(\vm)\Omega
\; \mbox{ with } \; \alpha_{\vn}(\vn) = 1
\label{phiN}
\end{equation}
where the coefficients are given by
\begin{eqnarray}
\alpha_{\vn}(\vm) = \delta(\vm,\vn) + \sum_{s=1}^\infty \gamma^s
\sum_{j_1<k_1}\sum_{\nu_1=1}^\infty \nu_1 \cdots \qquad\qquad\qquad\qquad \nonu  
\qquad\qquad\qquad \times \cdots
\sum_{j_s<k_s}\sum_{\nu_s=1}^\infty \nu_s
\frac{\delta(\vm,\vn+\sum_{r=1}^s \nu_r\vE_{j_rk_r})}{\prod_{\ell=1}^s
[\cE_0(\vn+\sum_{r=1}^\ell \nu_r \vE_{j_r k_r})-\cE_0(\vn)]}
. \label{amun}
\end{eqnarray} 
Thus the eigenfunctions of the Sutherland model are
\begin{equation}
\Psi(\vx;\vn) = \langle \Omega,
\Phi_N(\vx)^* \tilde\eta_N(\vn)\rangle
\end{equation}
with $\vn$ satisfying Eq.\ \Ref{vn}, and the corresponding eigenvalues
are $\cE_0(\vn)$ in Eq.\ \Ref{cE0}.
\end{theorem}

\begin{proof} It is easy to see that the ansatz in Eq.\ \Ref{phiN} implies
\begin{equation}
\cH \eta_N(\vn) = \cE_\vn \eta_N(\vn)
\end{equation} 
if $\alpha_\vn(\vm)$ obeys
\begin{equation}
[\cE_0(\vm)-\cE_\vn ] \alpha_\vn(\vm) = \gamma\sum_{j<k}\sum_{\nu=1}^\infty
\nu \alpha_\vn(\vm -\nu \vE_{jk}).
\end{equation}
Demanding $\alpha_\vn(\vn)=1$ we get $\cE_\vn=\cE_0(\vn)$ (this is
true since the previous relation system of equations has triangular
structure in some natural sense).  We now make the ansatz
$$\alpha_\vn(\vm) = \sum_{s=0}^\infty \gamma^s \alpha_\vn^{(s)}(\vm),
\quad \alpha_\vn^{(0)}(\vm)=\delta(\vm,\vn),$$ and from the relation
above we get the recursion relations
$$\alpha_\vn^{(s)}(\vm) = \frac1{[\cE_0(\vm)-\cE_0(\vn)]} \sum_{j<k}
\sum_{\nu=1}^\infty \nu\alpha_\vn^{(s-1)}(\vm-\nu\vE_{jk}) .
$$ It is straightforward to solve this by induction and thus obtain
the result in Eq.\ \Ref{amun}.

To see that all the eigenstates are well-defined we note that
$$
m_j = (\vn +\vmu)_j = n_j -\sum_{k<j}\mu_{kj} + \sum_{k>j}\mu_{jk}
$$ 
where we use the notation
\begin{equation}
\vm-\vn= \vmu = \sum_{j<} \mu_{jk} \vE_{jk} , 
\end{equation} 
where it is important to note that only $\vm$'s with non-negative
integers $\mu_{jk}$ occur above. With that one can prove a highest
weight conditions showing that there are always only a finite number
of states $\hat\Phi_N(\vn+\vmu)\Omega$ different from zero (see
Appendix C.3 in \cite{CL}).  It is also important to note that
\begin{equation}
\cE_0(\vn+\vmu)-\cE_0(\vn) = \sum_{j=1}^N \left( 2\sum_{k=j+1}^N
\mu_{jk}[n_j-n_k + (k-j)\lambda] + \Bigl[ \sum_{k<j}\mu_{kj} -
\sum_{k>j}\mu_{jk}\Bigr]^2 \right),
\end{equation} 
which is a sum of positive terms if the condition in \Ref{vn} holds
true. This shows that there are no zero denominators. Moreover, the
sum in Eq.\ \Ref{amun} is always finite due to the Kronecker
delta. This shows that all $\eta_N(\vn)$ are finite linear combination
of states $\hat\Phi_N(\vm)\Omega$ and thus well-defined.
\end{proof}

It is straightforward to compute the functions
\begin{equation}
\hat F_N(\vx;\vn) \equiv \langle \Omega, \Phi_N(\vx)^*
\hat\Phi_N(\vn) \Omega \rangle\label{hatF}
\end{equation}
explicitly by taking the Fourier transform of the function
$F_{NN}(\vx;\vy)$ in Eq.\ \Ref{FNM} w.r.t.\ $\vy$ \cite{EL3}; see Eq.\
\Ref{hatF1} below.  In Ref.\ \cite{EL3} we also showed that
Theorem~\ref{thm1} reproduces Sutherland's solution \cite{Su}.

Note that Theorem~\ref{thm1} implies that the CFT operators
\begin{equation}
\tilde\Phi_N(\vn) = \sum_{\vm} \alpha_{\vn}(\vm) \hat\Phi_N(\vm)
\end{equation}
are eigenstates of $\cH$ in the following sense,
\begin{equation}
[\cH,\tilde\Phi_N(\vn)]\Omega = \cE_0(\vn) \tilde\Phi_N(\vn)\Omega .
\end{equation}
We thus can construct many more explicit formulas for eigenfunctions
of the Sutherland Hamiltonian $H_N$ as follows,
\begin{equation}
\langle \Omega, R^{N-M} \Phi_N(\vx)^*\tilde\Phi_M(\vn)^*
\Omega\rangle,
\end{equation}
where $N$ and $M$ can be different and the charge changing factor
$R^{N-M}$ is inserted to get a non-zero result. Thus there seems to
exist many different explicit formulas for each eigenfunction of the
Sutherland system.  We plan to give a more detailed discussion of
these formulas elsewhere.

\section{Anyons at finite temperature: elliptic case}
We now describe a generalization of the results in the previous
section to the elliptic case. We start with the heuristic argument
which led us to these results.

As is well-known, the fermion two-point correlation function at zero
temperature is $1/\theta(x-y)$ with $\theta(r)=\sin(r/2)$ (Eq.\
\Ref{2pt} for $\lambda=1$), and at finite temperature this
trigonometric function is replaced by
\begin{equation}
\theta(r) = \sin(\half r)\prod_{n=1}^\infty\Bigl( 1 - 2q^{2n}\cos r +
q^{4n}\Bigr) ,\quad q=\ee^{-\beta/2}\label{etet}
\end{equation} 
which is essentially the Jacobi Theta function
$\vartheta_1$,\,\footnote{~To be precise: $\theta(r) =
\vartheta_1(r/2)/[2 q^{1/4} \prod_{n=1}^\infty (1-q^{2n})]$
\cite{WW}.} with $\beta>0$ the inverse temperature; see e.g.\
\cite{CH} and references therein. As discussed in Section 2.1, at zero
temperature the function $\theta(r)=\sin(r/2)$ is the building block
of all anyon correlation functions, and in the derivation of Eq.\
\Ref{2nd} the interaction potential $V(r)$ arises from $\theta(r)$ as
follows \cite{CL},
\begin{equation}
V(r) = -\frac{\partial^2}{\partial r^2} \log \theta(r) . \label{3}
\end{equation}
It is not difficult to see that, if we insert $\theta(r)$ in
\Ref{etet}, $V(r)$ in Eq.\ \Ref{3} becomes identical with the elliptic
function in Eq.\ \Ref{eV}.  This suggested to us that one should be
able to generalize the results in the previous section to the elliptic
case by using anyons at finite temperature (which seem to be
equivalent to vertex operators on a torus; see e.g.\ \cite{B}).

\subsection{Finite temperature anyons and the eCS system}
To construct anyons at finite temperature one could use the
representation of anyons in Section 2.2, with the vacuum expectation
value replaced by the usual Gibbs state for the anyon Hamiltonian in
Eq.\ \Ref{aH},
\begin{equation}
\langle A\rangle_\beta \equiv \frac1\cZ {\rm Tr_0} \bigl( \ee^{-\beta 
H_0} A \bigr) \label{Gibbs}
\end{equation} 
for any operator $A$ in the Heisenberg algebra,\,\footnote{~Note that $A$
can be essentially any operator on the zero temperature anyon Hilbert
space.} with the usual normalization constant $\cZ={\rm Tr_0}
\exp(-\beta H_0)$. The trace ${\rm Tr}_0$ here is only over the charge
zero sector in the anyon Hilbert space (this turns out to be the
correct choice).

It is convenient to use an alternative construction based on a
reducible representation of the Heisenberg algebra obtained by the
well-known trick of doubling the degrees of freedom: we start with a
standard highest weight representation of two commuting copies of the
Heisenberg algebra,
\begin{equation}
[\hat\rho_A(n),\hat\rho_B(m)]=m\delta_{m,-n}\delta_{A,B},\quad
[\hat\rho_A(n),R_B] = \delta_{n,0}\delta_{A,B}R_A
\end{equation}
and 
\begin{equation}
R_A^*=R^{-1},\quad \hat\rho_A(n)^* = \hat\rho_A(-n), 
\end{equation}
for $A,B=1,2$ and $m,n\in\Z$, 
and with a highest weight vector $\Omega$ such that
\begin{equation}
\hat\rho_A(n)\Omega=0 \quad \forall n\geq 0,\quad A=1,2.
\end{equation}
Then 
\begin{eqnarray}&&
\pi(R)\equiv R_1,\qquad \pi(\hat\rho(0)) \equiv \hat\rho_1(0)
\nonumber \\[6pt]&&
\pi(\hat\rho(n))\equiv c_n\hat\rho_1(n) + s_n\hat\rho_2(-n) \quad
\forall n\neq 0
\end{eqnarray}
obviously defines a representation $\pi$ of the Heisenberg algebra,
provided that the parameters $c_n$ and $s_n$ obey the relations
\begin{eqnarray}
|c_n|^2 - |s_n|^2 = 1, \qquad \overline{c_n}=c_{-n},\qquad
\overline{s_n}=s_{-n} \label{scn} \qquad \forall n\neq 0\label{csn}
\end{eqnarray}
with the bar denoting complex conjugation. 

A particular choice for $c_n$ and $s_n$ allows us to represent the
Gibbs state as vacuum expectation value in this reducible
representation.

\begin{prop} For any element $A$ in the Heisenberg algebra, 
the vacuum expectation value in the representation $\pi$ above with
\begin{equation}
|c_n|^2 = \frac{1}{1-q^{2n}},\qquad |s_n|^2 = \frac{q^{2n}}{1-q^{2n}},
\qquad q=\ee^{-\beta/2} 
\label{cs1}
\end{equation}
is identical with the finite temperature expectation value of $A$,
\begin{equation}
\langle \Omega, \pi(A) \Omega\rangle = \langle A \rangle_\beta ,
\end{equation}
with the r.h.s.\ defined in Eq.\ \Ref{Gibbs}.
\end{prop}

To simplify notation we write in the following $A$ short for $\pi(A)$.

We then define finite {\it temperature anyons} as in Eq.\ \Ref{anyon},
with the definition of normal ordering changed in an obvious manner
(see Eq.\ (25) in \cite{EL2}).\,\footnote{~It is not necessary to change
the normal ordering prescription since it only amounts to a change by
finite constants, but it is convenient since this leads to somewhat
simpler formulas.}

The results in Sections 2.2 and 2.3 generalize in a surprisingly
straightforward manner:

\begin{prop}
\label{prop4} 
Propositions \ref{prop1} and \ref{prop2} hold true as they stands also
at finite temperature $\beta<\infty$, provided that the functions
$\theta(r)$ and $V(r)$ in Eqs.\ \Ref{tet} and \Ref{V} are replaced by
their elliptic generalizations in Eq.\ \Ref{etet} and \Ref{eV}.
\end{prop}

(For more details and proofs see \cite{EL2}.) 

However, the construction in Section 2.4 cannot be generalized
immediately due to a seemingly innocent detail: the second quantized
eCS Hamiltonian $\cH$ does {\it not} obey the highest weight condition
in Eq.\ \Ref{hw}.  To realize that this condition is crucial it is
helpful to note that Proposition \ref{prop4} is not restricted to the
finite temperature representations but holds true for {\it any}
irreducible representation $\pi$, i.e., $c_n$ and $s_n$ obeying
\Ref{csn} can be (essentially) arbitrary, and all results remain true
if we take
\begin{equation}
\theta(r)= \exp\Bigl(-\frac{\ii r}2 -\sum_{n=1}^\infty \frac1n\bigl[
  |c_n|^2 \ee^{\ii nr} + |s_n|^2 \ee^{-\ii nr}\bigr]
\Bigr)\label{tet1}
\end{equation} 
and
\begin{equation}
V(r) = -\sum_{n=1}^\infty n\bigl( |c_n|^2 \ee^{\ii nr} + |s_n|^2
\ee^{-\ii nr}\bigr) .\label{V1}
\end{equation}
This suggests that the integrable elliptic case must have a special
property. This property is as follows.

\begin{lemma}
\label{lemma3}
For the irreducible representation $\pi$ defined above and all
elements $A$ in the Heisenberg algebra, the following conditions holds
true:
\begin{equation}
\langle \Omega , [\cH, A] \Omega\rangle = 0 \label{hw2}
\end{equation}
if and only if $c_n$ and $s_n$ are as in Eq.\ \Ref{cs1} for some
$\beta>0$.
\end{lemma}

(The proof can be found in \cite{EL2}, Appendix B.)
 
It is interesting to note that the non-local part $\cC$ in the explicit
formula \Ref{cH} of the operator $\cH$ is representation dependent and
cannot be written in terms of the Heisenberg algebra alone; see
Eq.\ (58) in \cite{EL2}. Thus while it seems at first that the method
of doubling degrees of freedom to construct finite temperature
representations is only a convenient trick, we feel that there is more
to it.

As we now show, Eq.\ \Ref{hw2} is enough to construct the
eigenfunctions of the eCS system explicitly.

\subsection{Solution of the eCS system}
We insert $A=\Phi_N(\vx)^*\hat\Phi_N(\vn)$ into Eq.\ \Ref{hw2} and
obtain
$$ \langle [\cH, \Phi_N(\vx)] \Omega , \hat \Phi_N(\vn) \Omega\rangle
= \langle\Phi_N(\vx)\Omega,[\cH, \hat \Phi_N(\vn)] \Omega\rangle
$$ where we used the Leibniz rule for the commutator and that $\cH$ is
self-adjoint.

We thus see that, while Corollary \ref{corr1} is no longer true, we
can bypass it. All we need is the following generalization of
Proposition \ref{prop3}.

\begin{prop}
\label{prop5}
Proposition \ref{prop3} holds true also at finite temperature but with
Eq.\ \Ref{Heta} replaced by
\begin{equation}
[\cH,\hat\Phi_N(\vn)]\Omega = \cE_0(\vn) \hat\Phi_N(\vn)\Omega -\gamma
\sum_{\nu\in\Z} S_\nu \hat\Phi(\vn + \nu\vE_{jk} )\Omega
\end{equation}
where
\begin{equation}
S_0=0,\quad\ S_{\nu} = \frac{\nu}{1-q^{2\nu}}\quad \mbox{ and } \quad
S_{-\nu} = \frac{\nu q^{2\nu}}{1-q^{2\nu}}\quad\ \forall \nu>0
\label{Snu}
\end{equation} 
and $\cE_0(\vn)$ is as in Eq.\ \Ref{cE0}. 
\end{prop}
(The proof is as in the zero temperature case, only now we use Eq.\
\Ref{V1} and \Ref{cs1} for taking into account the interaction term
\cite{EL2}.)

With that it is possible to generalize the arguments leading to the
solution of the Sutherland system in Theorem \ref{thm1}. This provides
an alternative route to the explicit solution of the eCS system
announced in \cite{EL4}.

\begin{theorem} 
\label{thm2} 
Let
\begin{equation}
\hat F(\vx;\vn) \equiv \cP(\vx;\vn) \Psi_0(\vx) \label{hatF1} ,
\end{equation}
with 
\begin{equation}
\cP(\vx;\vn) \equiv \prod_{j=1}^N\Bigl[ \oint_{\cC_j} \frac{\dd
\xi_j}{2\pi\xi_j} \xi_j^{n_j} \Bigr] \frac{\prod_{1\leq j<k\leq
N}\Theta(\xi_j/\xi_k)^\lambda} {\prod_{j,k=1}^N\Theta(\ee^{\ii
x_j}/\xi_k)^\lambda}   \label{cP}
\end{equation} 
where
\begin{equation} 
\Theta(\xi) \equiv (1-\xi)\prod_{m=1}^\infty [(1-q^{2m}\xi)(1-q^{2m}/\xi)]
\end{equation} 
and the integration contours are nested circles in the complex plane
enclosing the unit circle,
\begin{equation}
\cC_j:\quad \xi_j=\ee^{\eps j} \ee^{\ii y_j},\quad -\pi\leq y_j\leq \pi,
\quad 0<\eps<\beta/N ,
\end{equation}
and
\begin{equation}
\Psi_0(\vx) \equiv \ee^{\ii N \lambda \sum_{j=1}^N x_j/2} \prod_{1\leq
j<k\leq N} \theta(x_j-x_k)^\lambda \label{to_rest}
\end{equation}
with the function $\theta(r)$ in Eq.\ \Ref{etet}.  Then for each
$\vn\in\Z^N$ satisfying the condition in \Ref{vn}, there is an
eigenfunction $\Psi(\vx;\vn)$ of the eCS Hamiltonian in Eqs.\
\Ref{H}, \Ref{eV} given by
\begin{equation}
\Psi(\vx;\vn) = \sum_{\vm} \alpha_{\vn}(\vm)\hat F(\vx;\vm)
\end{equation}
with the coefficients
\begin{eqnarray}
\alpha_{\vn}(\vm) = \delta(\vm,\vn) + \sum_{s=1}^\infty \gamma^s
\sum_{j_1<k_1}\sum_{\nu_1\in\Z} S_{\nu_1} \cdots \qquad\qquad\qquad\qquad
\nonu\qquad\qquad\qquad \times \cdots 
\sum_{j_s<k_s}\sum_{\nu_s\in\Z} S_{\nu_s}
\frac{\delta(\vm,\vn+\sum_{r=1}^s\nu_r\vE_{j_r k_r} )}{\prod_{r=1}^{s}
[\![\cE_0(\vn+\sum_{\ell=1}^r\nu_\ell\vE_{j_\ell k_\ell} )- \cE_\vn]\!]
_{\vn}^{\phantom s}}
\end{eqnarray}
with $\cE_0(\vn)$ in Eq.\ \Ref{cE0}, $S_\nu$ in Eq.\ \Ref{Snu}, and
$\cE_\vn$ the corresponding eigenvalue determined by the following
implicit equation:
\begin{eqnarray}
\cE_\vn = \cE_0(\vn) - \sum_{s=2}^\infty \gamma^s
\sum_{j_1<k_1}\sum_{\nu_1\in\Z} S_{\nu_1} \cdots
 \qquad\qquad\qquad\qquad\qquad\quad
\nonu\qquad\qquad\qquad \times \cdots
\sum_{j_s<k_s}\sum_{\nu_s\in\Z} S_{\nu_s}
\frac{\delta(\vzero,\sum_{r=1}^s\nu_r\vE_{j_r k_r} )} {
\prod_{r=1}^{s} [\![\cE_0(\vn+\sum_{\ell=1}^r\nu_\ell\vE_{j_\ell k_\ell}
)- \cE_\vn]\!]^{\phantom s}_{\vn}} .
\label{cE}
\end{eqnarray} 
The double square brackets above are to indicate that all terms which
are undefined for $\cE_\vn=\cE_0(\vn)$, i.e.\ where
$\sum_{\ell=1}^r\nu_\ell\vE_{j_\ell k_\ell}=\vzero$, should be
discarded,
\begin{eqnarray}
\frac1{[\![\cE_0(\vm) -\cE_{\vn}]\!]^{\phantom s}_{\vn}} \equiv
[1-\delta(\vn,\vm)]\frac1{[\cE_0(\vm) -\cE_{\vn}]} .
\end{eqnarray}
\end{theorem}

For simplicity we ignore the problem of so-called {\it resonances}
here; see \cite{EL5}. The generalization of our argument above to
prove this result is not completely obvious. An outline of a somewhat
empirical derivation of this result appeared in \cite{EL4}, and we
plan to give a self-contained proof in a future revision of Ref.\
\cite{EL5}.

To get eigenfunctions in one-to-one correspondence with the Sutherland
case one should impose the restriction in Eq.\ \Ref{vn}, even though
this restriction does not seem necessary in the derivation of this
result. However, we suspect that it is necessary to ensure that Eq.\
\Ref{cE} has a well-defined solution. 

{}From Eq.\ \Ref{cE} one can obtain by straightforward computations a
fully explicit formula for the eigenvalues as follows,
\begin{eqnarray} 
\cE_\vn = \cE_0(\vn) + \sum_{n=1}^\infty (-1)^n
\sum_{k_0,\cdots,k_{n-1}=0}^\infty 
\delta(n-1,\smalleq{\sum_{j=1}^{n-1} jk_j})
\nonu\times  
\delta(n,\smalleq{\sum_{j=1}^{n-1} k_j }) \binom{n-1}{k_0,\ldots,k_{n-1} } 
\prod_{j=0}^{n-1} [G_j(\vn)]^{k_j} 
\end{eqnarray} 
with
\begin{eqnarray}
G_k(\vn) \equiv \sum_{s=2}^\infty \gamma^{s}
\sum_{j_1<k_1}\sum_{\nu_1\in\Z} S_{\nu_1} \cdots
\sum_{j_s<k_s}\sum_{\nu_s\in\Z} S_{\nu_s}
\sum_{\ell_1,\ldots,\ell_s=0}^\infty \delta(k,\smalleq{\sum_{r=1}^s
\ell_r}) \nonu\times \frac{\delta(\smalleq{\sum_{r=1}^{s}\nu_r\vE_{j_r
k_r},\vzero)} }{ \prod_{r=1}^{s-1}
[\![\cE_0(\vn+\sum_{\ell=1}^{r}\nu_\ell\vE_{j_\ell k_\ell})
-\cE_0(\vn)]\!]_{\vn}^{1+\ell_s} },
\end{eqnarray}
and there are similarly explicit formulas for the coefficients
$\alpha_\vn(\vm)$ (we plan to give the latter in a revised version of
Ref.\ \cite{EL5}). All results stated above hold true at least in the
sense of formal power series in $q^2$, but there exist results which
seem to imply that all our series have a finite radius of convergence
in $q^2$ \cite{KT}.

Note that the functions $\cP(\vx;\vn)$ are symmetric in the variables
$z_j=\ee^{\ii x_j}$, and they can be expanded as Laurent series. For
$q=0$ they become symmetric polynomials which are, in fact, equal to
the so-called Jack polynomials \cite{McD,St}; see \cite{EL3}.  It is
also interesting to note that $\Psi_0(\vx)$ for $q=0$ is the
well-known ground state wave function of the Sutherland model
\cite{Su}, but it is {\it not} an eigenfunction of the eCS Hamiltonian
for $q>0$. We thus see that $q>0$ makes the solution much more
involved, even though it is still possible to write it down
explicitly.

\section{Final remarks} 

\noindent {\bf 1.} While we found the solution method for the eCS as
described in this paper, the final result can be proven without using
CFT: Theorem \ref{thm2} can also be derived from the functional
identity
\begin{equation}
H_N(\vx)F_{N,N}(\vx;\vy) = H_N(\vy)F_{N,N}(\vx;\vy) \label{id}
\end{equation}
for the anyon correlation function defined in Eqs.\ \Ref{FNM} and
\Ref{etet} and the eCS Hamiltonians in Eqs.\ \Ref{H} and \Ref{eV} but
acting on different arguments $\vx$ and $\vy$, as indicated. While
Eq.\ \Ref{id} is a simple consequence of Proposition \ref{prop4} and
Lemma \ref{lemma3}, it can also be proven by brute-force computations
using a well-known functional relation of the Weierstrass elliptic
functions \cite{EL5}. While this method of proof is elementary it
seems ad-hoc, and it would have been difficult for us to get the
details right without knowing the CFT results discussed in this paper.
In \cite{EL6} we were able to find many more such identities using
these CFT results.

\bigskip

\noindent {\bf 2.} The solution method based on a remarkable identity
as in \Ref{id} works for many more quantum integrable systems than the
ones discussed here.\,\footnote{~M.\ Halln\"as and E.\ Langmann, in
preparation.}

\bigskip

\noindent {\bf 3.} The motivation for starting this project was the
FQHE, and even though it seems now that our results are mainly of
interest in the context of integrable systems, we still hope that they
will be eventually relevant for the FQHE. We now shortly describe some
ideas in this direction.

As indicated in the introduction, we believe that the second quantized
Sutherland Hamiltonian $\cH$ for odd integers $\lambda$ describe the
FQHE system at filling $1/\lambda$: starting with a realistic
many-body Hamiltonian for the FQHE describing fermions in two
dimensions in a strong magnetic field and interacting via the Coulomb
potential, one can map this to a 1D interaction fermions system by
projection into the lowest Landau level; we believe that $\cH$
describes a fixed point of the renormalization group applied to this
latter Hamiltonian.  If one could substantiate this, the exact
eigenstates of $\cH$ computed in Section~2.3 would become relevant in
the theory of the FQHE. It would be also interesting to find
corresponding Hamiltonians describing more complicated filling
fractions and corresponding to more complicated ground states of
composite fermions \cite{Jain} --- perhaps these would also be
interesting integrable systems.

\bigskip

\noindent {\bf 4.} As discussed after Proposition \ref{prop2}, it is
natural to conjecture that there exist self-adjoint operators $\cH_n$
on the anyon Hilbert space so that
\begin{equation}
[\cH_n,\phi(x)]\Omega = \ii^n \frac{\partial^n}{\partial x^n} \phi(x)
\Omega
\end{equation}
for all non-negative integers $n$. In 1999 we\,\footnote{~E.~Langmann,
unpublished.} constructed an operator valued generating functional for
these operators by generalizing the method in \cite{CL} to all orders.
This functional is defined as follows,
\begin{equation}
\cW(a) = \sum_{n=0}^\infty \frac{(-\ii a)^n}{n!} \cH_n
\end{equation}
and thus obeys
\begin{equation}
[\cW(a),\phi(x)]\Omega=\phi(x+a)\Omega
\end{equation}
as a formal power series in $a$. We found
\begin{equation}
\cW(a) = \sum_{s=0}^\infty \frac{ \ii w_s(a) }{2 [\cos(\half
a)]^\lambda \tan(\half a)\lambda} \tilde\cW_s(a) \label{cW}
\end{equation} 
with 
\begin{equation} 
\tilde\cW_s(a) \equiv \int_{-\pi}^\pi \frac{\dd x}{2\pi} \bigl[
\cV_-(x;a) \frac{\partial^s}{\partial x^s} \cV_+(x;a) -I\delta_{s,0}
\Bigr],
\label{tcW}
\end{equation}
where 
\begin{equation}
\cV_\pm(x;a) \equiv \exp\Bigl( -\ii \sqrt{\lambda} \sum_{k=0}^\infty
\frac{a^{k+1}}{(k+1)!}\rho^{(k)}_\pm(x) \Bigr) 
\end{equation}
and
\begin{equation}
\rho_\pm (x) \equiv \half \sqrt{\lambda} Q + \sum_{n\neq 0} \hat\rho(\mp n)
\ee^{\mp \ii nx}
\end{equation}
are the creation and annihilation parts of the boson fields,
$\rho_\pm^{(k)}(x) = \partial^k\rho_\pm(x)/\partial x^k$, and the
coefficients $w_s(a)$ are determined by the recursion relations
\begin{equation}
w_0(a)\equiv 1,\qquad w_s(a) \equiv -\sum_{k=0}^{s-1} v_{s-k}(a)
w_k(a)
\end{equation}
where
\begin{equation}
v_k(a)\equiv \sum_{\ell=k}^\infty \frac1\lambda
\binom{\lambda}{\ell+1} [-\tan(\half a)]^{\ell} \left. \frac1{k!\ell!}
\frac{\dd^{\ell}}{\dd c^{\ell}}
\frac{[2\arctan(c)]^k}{(1+c^2)}\right|_{c=0}.
\end{equation}
Note that the operator $\tilde\cW_0(a)$ is equal to
$\int_{-\pi}^\pi\dd x\, \xx\!\phi(x+a)\phi(x)^*\!\xx$ and gives rise to the
local part of the $\cH_n$, and the $\tilde\cW_{s>0}(a)$ give non-local
corrections. It is straightforward to write a MATLAB
program which computes all $\cH_n$ explicitly (this is possible since
$w_s(a)=O(a^s)$, and thus only the operators $\tilde \cW_s(a)$ for $s<n$
contribute to $\cH_n$.)  One thus can check that $\cH_0=Q$,
$\cH_1=H_0$, $\cH_2=\cH$ as given in Section~2.2, and the next
operator is
\begin{eqnarray}
\cH_3 = \int_{-\pi}^\pi \frac{\dd x}{2\pi} \Bigl( \xx \Bigl[
\frac{\lambda}4\rho(x)^4 + \frac14 (\rho'(x))^2 - \frac18(3\lambda
-2)\rho(x)^2\Bigr]\xx \nonu 
+ \frac{3\ii}2 \sqrt{\lambda}(\lambda-1)
\bigl[\rho_-(x)(\rho_+(x)^2)' + \rho_-(x)^2\rho_+(x)'\bigr] \nonu 
- \frac12(2\lambda-1)(\lambda-1) \rho_-(x)\rho_+(x)'' 
\Bigr) 
\end{eqnarray} 
where the terms in the first line are local and the others are
non-local corrections.

It is easy to show that all $\cH_n$ are self-adjoint and annihilate
the vacuum $\Omega$.  It is natural to expect that these operators
$\cH_n$ provide a second quantization of the higher conservation laws
$H_N^{(n)}$ of the Sutherland system, but we have not been able to
prove this result.

\medskip

 \def\adma  {Adv.\,Math.}
 \def\comp  {Com\-mun.\,Math.\,Phys.}
 \def\Comp  {Com\-mun.\linebreak[0]Math.\,Phys.}
 \def\cpma  {Com\-pos.\,Math.}
 \def\duke  {Duke\,Math.\,J.}
 \def\imrn  {Int.\,Math.\,Res.\,No\-ti\-ces}
 \def\jofa  {J.\,Funct.\,Anal.}
 \def\jomp  {J.\,Math.\,Phys.}
 \def\jopa  {J.\,Phys.\,A:\,Math.\,Gen.}
 \def\lemp  {Lett.\,Math.\,Phys.}
 \def\Lemp  {Lett.\,Math.\linebreak[0]Phys.}
 \def\nupb  {Nucl.\,Phys.~B}
 \def\Nupb  {Nucl.\linebreak[0]Phys.~B}
 \def\phla  {Phys.\,Lett.~A} 
 \def\phlb  {Phys.\,Lett.~B}
 \def\phra  {Phys.\,Rev.~A}
 \def\phrb  {Phys.\,Rev.~B}
 \def\phrl  {Phys.\,Rev.\,Lett.}
 \def\Phrl  {Phys.\linebreak[0]Rev.\,Lett.}
 \def\ptps  {Progr.\,The\-or.\,Phys.\,Suppl.}

\bibliographystyle{amsalpha}

\begin{thebibliography}{AMOS95}

\bibitem[AJL97]{AJL}
J.~Avan, A.~Jevicki, and J.~Lee, \emph{Yangian-invariant field theory of
  matrix-vector models}, \nupb\ {486} (1997), 650--672

\bibitem[AMOS95]{AMOS}
H.~Awata, Y.~Matsuo, S.~Odake, and J.~Shiraishi, \emph{Collective field theory,
  {C}alogero-{S}utherland model and generalized matrix models}, 
  \phlb\ {347} (1995), 49--55

\bibitem[B88]{B}
D.~Bernard, \emph{On the {W}ess-{Z}umino-{W}itten models on the torus}, 
  \nupb\ {303} (1988), 77--93

\bibitem[BPS94]{BPS}
D.~Bernard, V.~Pasquier, and D.~Serban, \emph{Spinons in conformal field
  theory}, \Nupb\ {428} (1994), 612--628

\bibitem[C71]{C}
F.~Calogero, \emph{Solution of the one-dimensional {N} body problems with
  quadratic and/or inversely quadratic pair potentials}, 
  \jomp\ {12} (1971), 419--436

\bibitem[CH87]{CH}
A.L.~Carey and K.C.~Hannabuss, \emph{Temperature states on the loop groups,
  theta functions and the {L}uttinger model}, \jofa\ {75} (1987), 128--160

\bibitem[CL99]{CL}
A.L.~Carey and E.~Langmann, \emph{Loop groups, anyons and the
  {C}alogero-{S}utherland model}, \comp\ {201} (1999), 1--34

\bibitem[DI93]{DI}
J.~Dittrich and V.I.~Inozemtsev, \emph{On the structure of eigenvectors of the
  multidimensional {L}am\'e operator}, \jopa\ {26} (1993), L753--L756

\bibitem[EFK95]{EFK}
P.I.~Etingof, I.B.~Frenkel, and A.A.~Kirillov, \emph{Spherical functions on
  affine {L}ie groups}, \duke\ {80} (1995), 59--90

\bibitem[EK94]{EK}
P.I.~Etingof and A.A.~Kirillov, \emph{Representation of affine {L}ie algebras,
  parabolic differential equations and {L}am\'e functions}, 
  \duke\ {74} (1994), 585--614

\bibitem[FNP03]{P}
W.~Garc\'{\i}a Fuertes, J.~Fern\'{a}ndez N\'{u}\~{n}ez, and A.M.~Perelomov,
  \emph{A perturbative approach to the quantum elliptic {C}alogero-{S}utherland
  model}, \phla\ {307} (2003), 233--238

\bibitem[FV95a]{FV1}
G.~Felder and A.~Varchenko, \emph{Integral representation of solutions of the
  elliptic {K}nizhnik-{Z}amolodchikov-{B}ernard equations}, 
  \imrn\ {5} (1995), 221--233

\bibitem[FV95b]{FV2}
\bysame, \emph{Three formulas for eigenfunctions of integrable
  {S}chr{\"o}dinger operators}, \cpma\ 107 (1997), 143--175

\bibitem[H94]{Ha}
Z.N.C.~Ha, \emph{Exact dynamical correlation functions of Calogero-Sutherland model 
and one-dimensional fractional statistics}, \phrl\ {73} (1994), 1574--1577 

\bibitem[I95]{Iso}
S.~Iso, \emph{Anyon basis of c = 1 conformal field theory}, 
  \nupb\ {443} (1995), 581--595

\bibitem[J89]{Jain}
J.K.~Jain, \emph{Composite fermion approach for the fractional quantum {H}all
  effect}, \Phrl\ {63} (1989), 199--202

\bibitem[K98]{Kac}
V.~Kac, \emph{Vertex Algebras for Beginners}, second ed., University Lecture
  Series, vol.~10, American Mathematical Society, Providence 1998

\bibitem[KT02]{KT}
Y.~Komori and K.~Takemura, \emph{The perturbation of the quantum
  {C}alogero-{M}oser-{S}ut\-her\-land system and related results}, 
  \comp\ {227} (2002), 93--118

\bibitem[L00]{EL1}
E.~Langmann, \emph{Anyons and the elliptic {C}alogero-{S}utherland model},
  \lemp\ {54} (2000), 279--289

\bibitem[L01]{EL3}
\bysame, \emph{Algorithms to solve the (quantum) {S}utherland model}, 
  \jomp\ {42} (2001), no.~9, 4148--4157

\bibitem[L04a]{EL5}
\bysame, \emph{A perturbative algorithm to solve the (quantum) elliptic
  {C}alogero-{S}utherland model}, Preprint math-ph/0401029.

\bibitem[L04b]{EL2}
\bysame, \emph{Second quantization of the elliptic {C}alogero-{S}utherland
  model}, \Comp\ {247} (2004), 321--351

\bibitem[L04c]{EL6}
\bysame, \emph{Further remarkable identities related to the (quantum)
elliptic Calogero-Sutherland model}, Preprint math-ph/0406061

\bibitem[L04d]{EL4}
\bysame, \emph{An explicit solution of the (quantum) elliptic {C}alogero-%
  {S}utherland model}, Preprint math-ph/0407050

\bibitem[Lgh83]{La}
R.B.~Laughlin, \emph{Anomalous quantum {H}all effect: An incompressible quantum
  fluid with fractionally charged excitations}, \phrl\ {50} (1983), 1395--1398

\bibitem[McD79]{McD}
I.G.~Macdonald, \emph{Symmetric Functions and {H}all polynomials}, The
  Clarendon Press, Oxford University Press, New York 1979

\bibitem[MP94]{MP}
J.A.~Minahan and A.P.~Polychronakos, \emph{Density correlation functions in
  {C}alogero-{S}utherland models}, \phrb\ {50} (1994), 4236--4239

\bibitem[MS96]{MS}
V.~Marotta and A.~Sciarrino, \emph{From vertex operators to
  {C}alogero-{S}utherland models}, \nupb\ {476} (1996), 351--373

\bibitem[OP77]{OP}
M.A.~Olshanetsky and A.M.~Perelomov, \emph{Quantum completely integrable
  systems connected with semisimple {L}ie algebras}, \lemp\ {2} (1977), 7--13

\bibitem[PS86]{PS}
A.~Pressley and G.~Segal, \emph{Loop Groups},
  The Clarendon Press, Oxford University Press, New York 1986

\bibitem[PS01]{PaS}
V.~Pasquier and D.~Serban, \emph{Conformal field theory and edge excitations
  for the principal series of quantum {H}all fluids}, \phrb\ {63}
  (2001), 153311(4)

\bibitem[Se81]{Se}
G.~Segal, \emph{Unitary representations of some infinite-dimensional groups},
  \Comp\ {80} (1981), 301--342

\bibitem[Sk95]{S}
E.K.~Sklyanin, \emph{Separation of variables -- new trends}, 
  \ptps\ {118} (1995), 35--60

\bibitem[St89]{St}
R.P.~Stanley, \emph{Some combinatorial properties of {J}ack symmetric
  functions}, \adma\ {77} (1989), 76--115

\bibitem[Su72]{Su} 
B.~Sutherland, \emph{Exact results for a quantum
many body problem in one-dimension. 2}, \phra\ {5} (1972), 1372--1376

\bibitem[T00]{T}
K.~Takemura, \emph{On the eigenstates of the elliptic {C}alogero-{M}oser
  model}, \Lemp\ {53} (2000), 181--194

\bibitem[W90]{Wen}
X.G.~Wen, \emph{Chiral {L}uttinger liquid and the edge excitations in the
  fractional quantum {H}all states}, \phrb\ {41} (1990), 12838--12844

\bibitem[WW62]{WW}
E.T.~Whittaker and G.N.~Watson, \emph{A Course of Modern Analysis}, Fourth
  edition, reprinted, Cambridge University Press, New York 1962

\end{thebibliography}

\medskip 

\end{document}